\documentclass[
    ,final            
  ]
  {aipproc}

\layoutstyle{6x9}

\usepackage{graphicx} 
\usepackage{psfrag} 
\usepackage{subfigure} 
\usepackage{array} 
\usepackage{booktabs} 
\usepackage{amsfonts} 
\usepackage{amssymb} 
\usepackage{amsmath} 
\usepackage{latexsym} 
\usepackage{amsthm} 
\usepackage{slashed} 
\usepackage{bm} 
\usepackage{mathrsfs} 
\usepackage[usenames,dvipsnames]{color} 
\usepackage{subfigure}
\usepackage{color}

\newcommand{\be}{\begin{equation}}
\newcommand{\ee}{\end{equation}}

\begin{document}

\title{First implications of LHCb data \\ on models beyond the Standard Model}



\author{Maria Valentina Carlucci}{
  address={Physik-Department, Technische Universit\"at M\"unchen, \\ James-Franck-Stra{\ss}e, D-85748 Garching, Germany}
}

\begin{abstract}
We discuss the theoretical and experimental details of two of the main results obtained by LHCb with the 2011 data, namely the measurement of the mixing-induced CP-violation in the decay $B^0_{s} \rightarrow J/\psi \; \phi$ and the upper limits on the decays $B^0_{(s)} \rightarrow \mu^+ \mu^-$. Then we describe the possible strategies to obtain new constraints on two different New Physics models in the light of these results.
\end{abstract}

\maketitle


\section{Introduction}

In 2007 a discrepancy larger that 3$\sigma$ was found between the theoretical prediction of the phase of the $B_s$ mixing amplitude in the Standard Model (SM) and the tagged analyses of the decay $B_s \rightarrow J/\psi \; \phi$ by the CDF and D0 collaborations \cite{Bona:2008jn}. It was the first of several other tensions of 1-3$\sigma$ that arose in the analysis of different flavour observables \cite{Buras:2010wr}, calling for more precise data in order to finally establish if some New Physics (NP) had shown up. In this sense, the year 2011 and the first months of 2012 have been an exciting period for the flavour community, because of the release of both the last results from Tevatron and the first precision measurements from LHC, in particular LHCb.

The most recent results have put an end to the enthusiasm about two of the observables that were considered most promising in revealing the presence of NP. In fact, the previously cited measurement of the CP-violating phase in the decay $B_s^0 \rightarrow J/\psi \; \phi$ has been found to be well consistent with the SM within current uncertainties \cite{LHCb:2011aa}. Moreover, the new constraints on the decays $B^0_{(s)} \rightarrow \mu^+ \mu^-$, of great interest because of their theoretical clearness and their large NP allowance, are now very close to the SM prediction \cite{Aaij:2012ac}. In the second section of this contribution we will review the theoretical and phenomenological aspects of these two superstars of the flavour physics of the last years, and we will describe the details of the LHCb results.

A considerable part of the work in the analyses of the models Beyond the Standard Model (BSM) performed during the last years has been devoted on one side to check their consistency with the strict constraints coming from precision flavour measurements, and on the other side to understand if they were able to explain the tensions present in flavour observables; as a consequence, models predicting new sources of flavour violation in the $B_s^0$ system and models with room for enhanced branching ratios for $B^0_{(s)} \rightarrow \mu^+ \mu^-$ have become very popular \cite{Buras:2012ts}. The new LHCb data represent now a strict test for these models; in the third section of this contribution we will outline how two different BSM scenarios, with different flavour violation patterns, can be affected by these experimental results, waiting for more detailed studies that will be able to constrain more precisely their parameter space.

\section{Selected results from LHCb}

It is well known that Flavour Changing Neutral Currents (FCNCs), being strongly suppressed in the SM because of its peculiar flavour structure, are one of the most interesting processes in order to detect possible NP contributions. In particular, two types of FCNC processes can be considered: the oscillation of neutral mesons, and the classes of the so-called rare decays, i.e. the FCNC decays that are mediated by electroweak box and penguin type diagrams in the SM. The two observables that we are going to analyze are examples respectively of the first and of the second category.

\subsection{The mixing-induced CP-violation in the decay $B^0_{s} \rightarrow J/\psi \; \phi$}

The neutral meson systems $B_q - \bar{B}_q$ $(q=d,s)$ can be described by an effective non-hermitian Hamiltonian:
\be
\mathcal{H}_q = \begin{pmatrix} M_1^q & M_{12}^q \\ M_{12}^{q*} & M_1^{q} \end{pmatrix} - \frac{i}{2} \begin{pmatrix} \Gamma_1^q & \Gamma_{12}^q \\ \Gamma_{12}^{q*} & \Gamma_1^{q} \end{pmatrix} ~,
\ee
whose diagonalization leads to the mass and width eigenstates $| B_q^{L,H} \rangle = p | B_q \rangle \pm q | \bar{B}_q \rangle$. The off-diagonal elements $M_{12}^q$ and $\Gamma_{12}^q$ are responsible for the mixing phenomena; their magnitudes are physical observables, and can be determined from measurements of respectively the mass difference $\Delta m_q$ and the decay width difference $\Delta \Gamma_q$ between the heavy and light mass eigenstates; their phase difference $\phi_q$ is also a physical observable, strictly related to CP violation and measurable trough the study of decay asymmetries.

In decays to a final state $f$ which is accessible to both $B_q$ and $\bar{B}_q$ mesons, one can introduce the key quantity $\lambda_f = (q/p) (\bar{A}_f / A_f)$, where $A_f$ and $\bar{A}_f$ are respectively the decay amplitudes of $B_q \rightarrow f$ and $\bar{B}_q \rightarrow f$, and can identify two types of CP violation: $\mathcal{A}_{\text{CP}}^{\text{dir}} \propto 1- \left| \lambda_f \right|^2$ and $\mathcal{A}_{\text{CP}}^{\text{mix}} \propto \text{Im} \lambda_f$. If $f$ is a CP eigenstate with eigenvalue $\eta_f$, then $\mathcal{A}_{\text{CP}}^{\text{dir}} \neq 0$ implies $|A_f| \neq |\bar{A}_f|$, meaning direct CP violation, while  $\mathcal{A}_{\text{CP}}^{\text{mix}}\neq 0$ measures mixing-induced CP violation in the interference between the two decays \cite{Dunietz:2000cr}. Moreover, if the decay is dominated by $b \rightarrow c \bar{c} s$ tree amplitude, one finds that the direct CP violation vanishes and that the time dependent asymmetry
\be
a_{\text{CP}}(t) \equiv \frac{\Gamma(\bar{B}_q(t) \rightarrow f) - \Gamma(B_q(t) \rightarrow f)}{ \Gamma(\bar{B}_q(t) \rightarrow f) + \Gamma(B_q(t) \rightarrow f)}
\ee
reduces to
\be
a_{\text{CP}}(t) = -\; \frac{ \eta_f \sin \phi_q \,  \sin(\Delta m_q \, t)}{\cosh \left( \Delta \Gamma_q \, t/2 \right) -  \eta_f |\cos \phi_q | \sinh \left( |\Delta \Gamma_q| \, t/2 \right)} ~,
\ee
giving direct access, once $\Delta m_q$ has been fixed, to $\Delta \Gamma_q$ and $\phi_q$. In addition, in these particular conditions the phases $\phi_q$ give directly the unitarity triangle angles, $\phi_d \approx 2 \beta$ and $\phi_s \approx - 2 \beta_s$, but, if there is NP in $M_{12}^q$ or in the decay amplitudes, the measured value of $\phi_q$ can differ from the true values of $\beta_{(s)}$ \cite{Bediaga:2012py}.

As it satisfies the previous conditions, and because of the clean experimental signature, the decay $B^0_{s} \rightarrow J/\psi \; \phi$ is considered the golden-plated mode for the measurement of this mixing-induced CP-violation for the $B_s - \bar{B}_s$ system. Since $J/\psi$ and $\phi$ are vector-mesons, the CP parity of the final state depends on the relative angular momentum, and an angular analysis is necessary in order to separate the CP parities.


After the already cited high-impact results from CDF and D0 in 2007, whose 2010/2011 updates showed however softened deviations, in the end of 2011 LHCb released the results of its first tagged analysis of $B^0_{s} \rightarrow J/\psi \; \phi$, using 0.34 fb$^{-1}$ of data \cite{LHCb:2011aa}. In the update with the full data sample of 1.0 fb$^{-1}$ collected before 2012 \cite{LHCb2012}, with approximately 21,200 flavour tagged $B_s^0 \rightarrow J/\psi (\rightarrow \mu^+ \mu^-) \;  K^+ \; K^-$ candidate events with $K^+ \; K^-$ invariant mass within $\pm$ 12 MeV of the $\phi$ mass, obtained $pp$ collisions $\sqrt{s} = 7$ TeV, they find:
\begin{subequations}
\be
\phi_s = -0.001 \pm 0.101 \text{(stat)} \pm 0.027 \text{(syst)} \; \text{rad} ~,
\ee
\be
\Delta \Gamma_s = 0.116 \pm 0.018 \text{(stat)} \pm 0.006 \text{(syst)} \; \text{ps}^{-1} ~,
\ee
\end{subequations}
that are the world's most precise measurement of $\phi_s$ and the first direct observation for a non-zero value for $\Delta \Gamma_s$. They are fully compatible with the SM predictions, whose updated values read \cite{Lenz:2011ti}
\be
\phi_s = 0.0038 \pm 0.0010 \; \text{rad} ~, \qquad \qquad \Delta \Gamma_s = 0.087 \pm 0.021 \; \text{ps}^{-1} ~.
\ee
LHCb has also published a paper \cite{Aaij:2012eq} which determined the sign of $\Delta \Gamma_s$ to be positive at 4.7$\sigma$ confidence level by exploiting the interference between the $K^+ \; K^-$ S-wave and P-wave amplitudes in the $\phi$(1020) mass region; this resolved the two-fold ambiguity in the value of $\phi_s$ for the first time.

\subsection{The decays $B^0_{(s)} \rightarrow \mu^+ \mu^-$}

These decays are a special case among the electroweak penguin processes, as they are chirality-suppressed in the SM and are most sensitive to scalar and pseudoscalar operators $O^{(}{}'{}^{)}_{S,P}$. In fact, the branching fraction can be expressed as
\begin{multline}
\mathcal{B} (B^0_q \rightarrow \mu^+ \mu^-) = \frac{G_F^2 \alpha^2}{64 \pi^3} f_{B_q} \tau_{B_q} m^3_{B_q} | V_{tb} V^{*}_{tq} |^2 \sqrt{1-\frac{4 m_{\mu}^2}{m^2_{B_q}}} \\
\times \left\{ \left( 1-\frac{4 m_{\mu}^2}{m^2_{B_q}} \right) | C_S - C'_S |^2 + \left| (C_P - C'_P) + 2 \frac{m_{\mu}}{m_{B_q}} (C_{10} -C'_{10}) \right|^2 \right\} ~,
\end{multline}
and within the SM $C_S$ and $C_P$ are negligibly small and the dominant contribution of $C_{10}$ is helicity suppressed. In the SM branching fraction the main source of uncertainty is caused by the $B_s^0$ decay constant $f_{B_s}$, but there has been significant progress in theoretical calculations of this quantity in recent years \cite{Laiho:2009eu}. The most recent predictions are \cite{Buras:2012ru}
\begin{subequations}
\be
\mathcal{B} (B^0_s \rightarrow \mu^+ \mu^-)_{\text{SM}} = (3.23 \pm 0.27) \times 10^{-9} ~,
\ee
\be
\mathcal{B} (B^0_d \rightarrow \mu^+ \mu^-)_{\text{SM}} = (1.07 \pm 0.10) \times 10^{-10} ~,
\ee
\end{subequations}
and it has been recently shown that a correction factor $r(\Delta \Gamma_s) = 0.91 \pm 0.01$ have to be considered when comparing $\mathcal{B} (B^0_s \rightarrow \mu^+ \mu^-)_{\text{SM}}$ with experiments \cite{DeBruyn:2012wk}.

A very interesting feature of these processes is that the coefficients $C_i$ are the same for $B_s^0$ and $B^0$ in any physics scenario that obeys Constrained Minimal Flavour Violation (CMFV), and hence the ratio $\mathcal{B} (B^0_s \rightarrow \mu^+ \mu^-) / \mathcal{B} (B^0 \rightarrow \mu^+ \mu^-)$ represents a very useful probe of this flavour pattern. In particular, the correlation between the two decays can be expressed with very small uncertainties \cite{Buras:2003jf}:
\be
\frac{\mathcal{B} (B^0_s \rightarrow \mu^+ \mu^-)}{\mathcal{B} (B^0 \rightarrow \mu^+ \mu^-)} = 
\frac{\tau(B_s)}{\tau(B_d)} \frac{m_{B_s}}{m_{Bs}} \frac{F^2_{B_d}}{F^2_{B_s}} \left| \frac{V_{td}}{V_{ts}} \right|^2 r(\mu^+ \mu^-) =
\frac{\hat{B}_d}{\hat{B}_s} \frac{\tau(B_s)}{\tau(B_d)} \frac{\Delta m_s}{\Delta m_d} r ~,
\ee
where $r(\mu^+ \mu^-) = r = 1$ in CMFV. This and other relations have been proposed as \emph{standard candles of flavour physics} \cite{Buras:2012ts} meaning that deviations from them may help in identifying the correct NP scenario.

Until last year, only high upper limits were available for both decays, leaving large space to many NP models, especially the ones with extended Higgs sector, that predict enhanced branching fractions. In 2011 CDF, CMS and LHCb published their new results, and at the beginning of 2012 LHCb set the world best limits \cite{Aaij:2012ac}: at 95\% C.L.,
\be
\mathcal{B} (B^0_s \rightarrow \mu^+ \mu^-) < 4.5 \times 10^{-9} ~, \qquad \qquad \mathcal{B} (B^0_d \rightarrow \mu^+ \mu^-) < 1.0 \times 10^{-9} ~;
\ee
that are now very close to the SM predictions.

\section{Impact on NP models}

Several phenomenological analyses of models BSM performed during the last years have been conducted keeping in mind the goal of being ready for the moment in which new and more precise experimental data would have been available; the so-called ``DNA tests'' of NP models \cite{Altmannshofer:2009ne}, as well as the correlation plots \cite{Straub:2010ih}, are examples of the tools developed for this purpose. For this reason, some qualitative or even semi-quantitative statements about the impact of the recent LHCb data on these models can be made before performing new complete numerical analyses. We will consider two NP models, with very different flavour patterns, showing that the new results put some of them in difficulty and some others back in the game.
\begin{itemize}
\item {\bf 2HDM$_{\overline{\text{MFV}}}$ \cite{Buras:2010mh}.} In presence of two Higgs doublets, the imposition of MFV is effective in suppressing FCNCs, and still the presence of flavour-blind phases is allowed. This model presented the very appealing characteristic that not only a large phase in the $B_s$ mixing was possible, but that it would have automatically solved the $S_{\psi \, K_S}-\epsilon_K$ tension. However, because of the small value for $\phi_s$ found by LHCb, this model not only loses this possibility, but is also put strongly under pressure.
\item {\bf Gauged Flavour Symmetries \cite{Grinstein:2010ve}.} Based on the assumption that the $SU(3)^3$ flavour symmetry of the SM is a gauge symmetry of Nature spontaneously broken by the vevs of new scalar fields, this model presents the elegant feature that the exotic fermion fields that need to be introduced in order to make the theory anomaly-free generate a mechanism of inverse hierarchy that suppresses FCNCs. We found that large corrections to $\epsilon_K$, $S_{\psi \, K_S}$ and $S_{\psi \; \phi}$ are allowed, but that requiring $\epsilon_K$ and $S_{\psi \, K_S}$ to be both in agreement with experiments permitted only small deviations from the SM value of $S_{\psi \; \phi}$ \cite{Buras:2011wi}. While at the time of our analysis this appeared as a possible problem for the model, this result is now fully consistent with the LHCb data.
\end{itemize}

%
%
%
%
%
%
%
%
%
%
%
%
%
%
%


\enlargethispage*{\baselineskip}
\begin{theacknowledgments}
I would like to thank Andrzej Buras for being such an inspiring and comprehensive PhD supervisor, Jennifer Girrbach for useful discussions, Pietro Colangelo and Fulvia De Fazio for manteining a kind and profitable relation, Marco Bardoscia for the invaluable support. This work has been supported in part by the Graduiertenkolleg GRK 1054 of DFG.
\end{theacknowledgments}

%
%
\bibliographystyle{aipproc}   
%
%
%


\end{document}